\documentstyle[prd,twocolumn,aps,epsf]{revtex}
\newcommand{\beq}{\begin{equation}}
\newcommand{\eeq}{\end{equation}}
\newcommand{\beqa}{\begin{eqnarray}}
\newcommand{\eeqa}{\end{eqnarray}}
\newcommand{\bea}{\begin{array}}
\newcommand{\ena}{\end{array}}
\newcommand{\singlefig}[2]{
\begin{center}
\begin{minipage}{#1}
\epsfxsize=#1
\epsffile{#2}
\end{minipage}
\end{center}}
\newcommand{\segmentfig}[3]{
\begin{minipage}{#1}
\epsfxsize=#1
\epsffile{#2}
\begin{center}
{\small \mbox{#3}}
\end{center}
\end{minipage}}
\newcommand{\lsim}{\mbox{\raisebox{-1.ex}{$\stackrel
     {\textstyle<}{\textstyle \sim}$}}}
\begin{document}
\draft
\title{Thermodynamic properties of massive dilaton black holes II}
\author{Takashi Tamaki\thanks{electronic mail
:tamaki@gravity.phys.waseda.ac.jp}}
\address{Department of Physics, Waseda University,
Ohkubo, Shinjuku, Tokyo 169-8555, Japan}
\date{\today}
\maketitle
\begin{abstract}
We numerically reanalyze static and spherically symmetric 
black hole solutions in an Einstein-Maxwell-dilaton system 
with a dilaton potential $m_{d}^{2}\phi^{2}$. We investigate 
thermodynamic properties for various dilaton coupling constants and 
find that thermodynamic properties change at a critical dilaton mass 
$m_{d,crit}$. For $m_{d}\geq m_{d,crit}$, the black hole becomes an 
extreme solution for a nonzero horizon radius $r_{h,ex}$ as 
the Reissner-Nordstr\"om black hole. However, if $m_{d}$ is nearly 
equal to $m_{d,crit}$, there appears a solution of smaller horizon 
radius than $r_{h,ex}$. For $m_{d}<m_{d,crit}$, a solution continues to 
exist until the horizon approaches zero. The Hawking temperature in the 
zero horizon limit resembles that of a massless dilaton black hole for 
arbitrary dilaton coupling constant. 
\end{abstract}

\section{Introduction}
The black hole solution in the Einstein-Maxwell-dilaton system found 
by Gibbons and Maeda, and independently by Garfinkle, Horowitz 
and Strominger (GM-GHS) has many distinctive properties from 
those of the conventional Reissner-Nordstr\"om (RN) black hole\cite{GM-GHS}. 
Among the most important are its thermodynamic properties. If we 
consider the evaporation process of the black hole, its final fate 
varies depending on the dilaton coupling. 

For dilaton couplings $\gamma <1$, the Hawking temperature approaches 
zero by reducing the mass, while it diverges for dilaton couplings $\gamma >1$ 
in the zero horizon limit\cite{foot2}. For the intermediate value $\gamma =1$, 
the temperature remains finite and nonzero in this limit. 

These facts tell us that the scalar field plays a role in 
determining the properties of black holes. It is worth mentioning that 
hairy black holes\cite{SU2EYMD,Kunz,Inada} which have a Yang-Mills field and a 
Higgs field also change qualitatively when scalar hair such as 
a Brans-Dicke scalar field is added\cite{Tamaki}. 

However, since the dilaton field has been predicted to have 
a TeV scale mass and a massless dilaton field is excluded by 
experimental considerations\cite{Will}, we need to investigate black hole 
solutions with a massive dilaton field as a more realistic case.  
Several authors have discussed this possibility and found some interesting 
properties\cite{HH}. For example, these black holes may have three horizons, or 
wormholes outside the event horizon in the string metric near 
the extreme solution for the magnetically charged solution. However, 
since the inclusion of a mass term makes it hard to obtain a solution in 
closed analytic form, there remains some room for further 
investigation\cite{foot}. 

We have investigated previously the thermodynamic properties of these black holes 
for the particular dilaton coupling $\gamma =1$\cite{TY}. However, the value of 
the dilaton coupling is not fixed experimentally. Moreover, theoretical 
considerations do not necessarily favor a dilaton coupling equal to $1$. 
For example, $\gamma =\sqrt{3}$ is suggested by a four-dimensional effective 
model obtained from the five-dimensional Kaluza-Klein theory. 
 
We therefore concentrate here on 
studying the dependence of the dilaton couplng and clarify some 
aspects which have not been investigated so far. 

This paper is organized as follows. In Sec. II, we describe our model, 
basic equations and boundary conditions. We briefly review the GM-GHS 
solution in Sec. III. Our main results are given in Sec. IV and V, 
with concluding remarks in Sec. VI. 
Throughout this paper, we use the units $G=c=\hbar =1$. 

\section{Basic Equations}

We take the action as follows, 
\beqa
S = \int
d^{4}x\sqrt{-g}\left[R-2(\nabla\phi)^{2}-e^{-2\gamma\phi}F^{2}
-2V(\phi) \right],
\label{eqn:action}
\eeqa
where $\phi$, $\gamma$ and $V(\phi)$ are a dilaton field, its coupling 
constant and its potential, respectively. 
The Einstein equations, the dilaton equation and the Maxwell equation are 
\beqa
G_{\mu\nu} =2\left(\nabla_{\mu}\phi\nabla_{\nu}\phi+e^{-2\gamma\phi}
F_{\mu}^{\lambda}F_{\nu\lambda}\right)   \nonumber \\
-g_{\mu\nu}\left[(\nabla\phi )^{2}
-\frac{1}{2}e^{-2\gamma\phi}
F^{\alpha\beta}F_{\alpha\beta}-V\right], 
\label{eqn:einstein0}  \\ 
\Box\phi+\frac{\gamma}{2}e^{-2\gamma\phi}F^{\mu\nu}F_{\mu\nu}
-\frac{1}{2}\frac{dV(\phi)}
{d\phi}=0 ,
\label{eqn:dilaton0}  \\
\nabla_{\mu}F^{\mu\nu}-2\gamma(\nabla_{\mu}\phi)F^{\mu\nu}=0.
\label{eqn:maxwell0}
\eeqa
We consider the static and spherically symmetric metric as 
\beqa
ds^{2}=-f(r) e^{-2\delta(r)}dt^{2}+f(r)^{-1}dr^{2}
+r^{2}d\Omega^{2},
\label{eqn:metric}
\eeqa
where $f(r):=1-2m(r)/r$. We choose a gauge potential 
$A_{\mu}$ as 
\beqa
A_{\mu}=(A(r),0,0,Q_{m}\cos\theta),
\label{eqn:gauge}
\eeqa
where $Q_{m}$ is a magnetic charge. We do not consider 
a dyonic black hole which has both an electric and a magnetic charge 
here. Below, we consider a dilaton potential $V=m_{d}^{2}\phi^{2}$ 
where $m_{d}$ is the dilaton mass, for simplicity. 
In this case, the boundary conditions at spatial infinity 
to satisfy asymptotic flatness are 
\begin{eqnarray}
m(\infty) =: M=const.,\ \ \delta (\infty)=0, \ \ 
\phi(\infty)=0. 
\label{atinf}
\end{eqnarray}
From the Maxwell equation, we obtain 
\begin{eqnarray}
(rK)'+K=0, 
\label{Max}
\end{eqnarray}
where $':=d/dr$ and $K:=A'e^{-2\gamma \phi +\delta}$. 
By integrating this equation, we obtain 
\begin{eqnarray}
r^{2}K=const.=:Q_{e}. 
\label{Max2}
\end{eqnarray}
$Q_{e}$ can be interpreted as electric charge because of the 
condition (\ref{atinf}). Finally, our basic equations are 
\beqa
&& m'=\frac{r^{2}}{2}\left(f\phi'^{2}+B+V\right), 
\label{eqn:einstein}   \\
&&\delta'=-r\phi'^{2},  \\    
&&A'=\frac{Q_{e}}{r^{2}}e^{2\gamma\phi-\delta}, 
\label{eqn:maxwell}    \\   
&&\phi''+\frac{2\phi'}{r}+   
\frac{1}{f}\left[
\phi'\left\{
\frac{2m}{r^2}-r\left(B+V\right)\right\}\right.  
\nonumber  \\ 
&&\left.+\gamma\left(\frac{Q_{m}^{2}}{r^{4}}e^{-2\gamma\phi}
-\frac{Q_{e}^{2}}{r^{4}}e^{2\gamma\phi}\right)
-\frac{1}{2}\frac{dV}{d\phi}\right]=0 ,
\label{eqn:dilaton}
\eeqa
where we use the abbreviation 
\beqa
B:=\frac{Q_{m}^{2}}{r^{4}}e^{-2\gamma\phi}
+\frac{Q_{e}^{2}}{r^{4}}e^{2\gamma\phi}\ . 
\eeqa
We also assume the existence of a regular 
event horizon at $r=r_{h}$. So we have 
\begin{equation}
m_{h}=\frac{r_h}{2},\;\; \delta_h< \infty ,
\;\;  \phi_h<\infty ,\;\;  
\label{mth}  
\end{equation}
\begin{equation}
\phi'_{h}=\frac{r(dV_{h}/d\phi -2\gamma B_{h})}{
2\left\{1-r_{h}^{2}(B_{h}+V_{h})\right\}}.
\label{ath}
\end{equation}
The variables with subscript $h$ are evaluated 
at the horizon. We will obtain the black hole 
solution numerically by determining $\phi_{h}$ 
iteratively to satisfy these conditions. 

\section{The GM-GHS solution}
We briefly describe the GM-GHS solution, i.e., the 
massless dilaton black hole, to compare with the massive dilaton 
black hole. To describe the exact solution, it is 
convenient to use the following spherically symmetric ansatz: 
\begin{eqnarray}
ds^{2}=-\lambda^{2} dt^{2}+
\frac{1}{\lambda^{2} }dr^{2}+R^{2}d\Omega^{2}\ , 
\label{forGM-GHS}
\end{eqnarray}
where $\lambda$ and $R$ are functions of $r$ only. 

We consider the electrically charged case. Since this 
system has an electric-magnetic duality $F \to e^{-2\phi}\tilde{F}$, 
$\phi \to -\phi$, we can easily obtain the results for the 
magnetically charged case. With the above ansatz, we obtain 
\begin{eqnarray}
e^{2\phi}&=& \left(1-\frac{r_{-}}{r}\right)^{2\gamma/(1+\gamma^{2})}  \ , 
\label{scalarGM}   \\ 
\lambda^{2}&=& \left(1-\frac{r_{+}}{r}\right)
\left(1-\frac{r_{-}}{r}\right)^{(1-\gamma^{2})/(1+\gamma^{2})} \ ,
\label{lambdaGM}   \\
R&=& r\left(1-\frac{r_{-}}{r}\right)^{\gamma^{2}/(1+\gamma^{2})}\ ,
\label{RGM}   
\end{eqnarray}
where we have set the asymptotic value of the dilaton field to zero. 
$r_{+}$ and $r_{-}$ are the event horizon and the inner horizon, 
respectively. They are related to the gravitational mass $M$ and 
charge $Q_{e}$ by 
\begin{eqnarray}
M&=&\frac{r_{+}}{2}+\left(\frac{1-\gamma^{2}}{1+\gamma^{2}}
\right)\frac{r_{-}}{2}\ ,
\label{massGM}   \\
Q_{e}&=&\left(\frac{r_{+}r_{-}}{1+\gamma^{2}}\right)^{1/2}\ .  
\label{magGM}
\end{eqnarray}
The above solution has many interesting features, one of 
which is that the inner horizon is singular and 
this singularity is spacelike for any nonzero 
value of $\gamma$. The second feature is thermodynamical. 
Temperature $T$ is written as 
\begin{eqnarray}
T=\frac{1}{4\pi r_{+}}\left( 1-\frac{r_{-}}{r_{+}}
\right)^{\frac{1-\gamma^{2}}{1+\gamma^{2}}  }\ . 
\end{eqnarray}
This vanishes in the zero horizon area limit for $\gamma <1$, 
whereas it has a finite value $1/(8\pi M)$ for $\gamma =1$ and 
diverges for $\gamma >1$ in the same limit. Thus, these differences 
are caused by the dilaton field and to examine this feature for 
the massive dilaton black hole is one of our basic motivations. 

\section{Properties of the massive dilaton black hole}
First, we summarize the results of previous papers which do not depend on 
the dilaton potential $V$ whenever $V$ is convex\cite{HH}: 
(i) The dilaton is monotonically increasing (decreasing) outside the 
horizon for the electrically (magnetically) charged solution. 
(ii) For $Q_{e}m_{d}\lsim 1$ (or $Q_{m}m_{d} \lsim 1$), there can be 
only one horizon. In previous papers only 
the case $\gamma =1$ is considered, but it is trivial to extend the above 
results to arbitrary $\gamma$. 

When $V$ is an even function, we can also obtain 
results for the magnetically charged case from the electrically 
charged case by means of the electric-magnetic duality mentioned above. 
Consider for example the magnetically charged 
solutions with $Q_{m}=0.1$. 

First, we consider the intrinsic difference of the dilaton field with 
a mass term from that of the massless dilaton field in an asymptotic region. 
For the GM-GHS solution, the dilaton field goes as 
$e^{-2\phi} \sim 1+2\Sigma /r$ in the asymptotic region, 
where $\Sigma$ is a dilaton charge. The latter is expressed as 
\begin{eqnarray}
\Sigma =-\frac{Q_{m}^{2}}{2M}\ . 
\end{eqnarray}
This is known as a secondary charge since it is completely 
determined by the gravitational mass and the magnetic charge. 

On the contrary, for the massive dilaton black hole, the dilaton field 
behaves as $\phi \sim \gamma Q_{m}^{2}/m_{d}^{2}r^{4}$\cite{HH} since 
Eq. (\ref{eqn:dilaton}) can be approximated as 
\begin{eqnarray}
\gamma\frac{Q_{m}^{2}}{r^{4}}=\frac{1}{2}\frac{dV}{d\phi} ,
\label{eqn:dilaton2}
\end{eqnarray}
by neglecting the derivative terms of $\phi$ in the asymptotic region. 
Thus, the dilaton charge is lost because of the dilaton potential. 

\begin{figure}[htbp]
\singlefig{10cm}{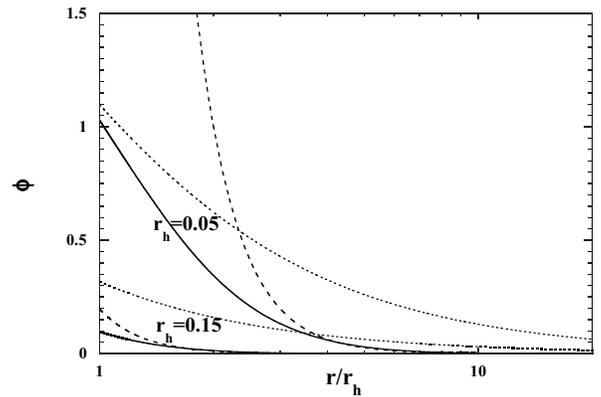}{}
\caption{The behavior of the dilaton field with $\gamma =1$ for 
$m_{d}=0$, $10$ and the approximate solution are shown with dotted lines, 
solid lines and dashed lines, respectively. 
\label{Fig1}}
\end{figure}

We show the behavior of the dilaton field outside the horizon 
for horizon radii $r_{h}=0.15$ and $0.05$, dilaton mass $m_{d}=10$ 
and dilaton coupling $\gamma =1$ in Fig.~\ref{Fig1} (solid lines). 
For comparison, we also show those of the GM-GHS solution and the 
approximate solution $\phi =\gamma Q_{m}^{2}/m_{d}^{2}r^{4}$ 
(dotted lines and dashed lines, respectively.). 

It is seen that the deviation from the approximate 
solution becomes large for $r_{h}=0.05$ compared with that for 
$r_{h}=0.15$. This is related to the fact that the massive dilaton field 
behaves like a massless one within its Compton wavelength scale 
$\sim 1/m_{d}$. 

On the other hand, since the influence of the dilaton 
field becomes negligible for $r\agt 1/m_{d}$, 
the solution is expected to approach the RN solution for 
large $m_{d}$ or large horizon radius. 
Thus, we may expect a qualitative difference depending on the scale of 
the horizon radius. 

This behavior may influence other structures. Actually, a wormhole 
structure outside the event horizon in the string metric near the extreme 
black hole of the magnetically charged case, which 
is absent for the GM-GHS solution, is reported in Ref. \cite{HH}. 

Next, we turn our attention to sequences of solutions, corresponding to 
various horizon radii. We analyze the conditions for the existence 
of a degenerate horizon along the same lines as Ref. \cite{HH}. If the 
horizon becomes degenerate at $r=r_{d}$, we have $m'=1/2$ 
at the horizon, which leads to 
\begin{eqnarray}
1=\frac{Q_{m}^{2}e^{-2\gamma\phi_{d}}}{r_{d}^{2}}+
m_{d}^{2}\phi_{d}^{2}r_{d}^{2},  \label{dege}  
\end{eqnarray}
from Eq.(\ref{eqn:einstein}). Note that we can rewrite Eq.(\ref{eqn:dilaton}) 
by using Eq.(\ref{eqn:einstein}) as 
\begin{eqnarray}
f\left(\phi ''+\frac{2\phi '}{r}\right)+\phi '(f'+rf\phi '^{2}) 
\nonumber \\
+\frac{\gamma Q_{m}^{2}}{r^{4}}e^{-2\gamma \phi}
-m_{d}^{2}\phi=0 .  \label{inter}  
\end{eqnarray}
Then, by using $f_{d}=f_{d}'=0$, we have 
\begin{eqnarray}
0&=&\frac{\gamma Q_{m}^{2}e^{-2\gamma\phi_{d}}}{r_{d}^{2}}-
m_{d}^{2}\phi_{d}r_{d}^{2}.  \label{dege2}
\end{eqnarray}
Note that the finiteness of $\phi_{d}$ and $\phi_{d}'$ at 
$r=r_{d}$ are assumed implicitly. From these conditions, we obtain 
\begin{eqnarray}
\frac{1}{\phi_{d}(\phi_{d}+1/\gamma)m_{d}^{2}}&=&r_{d}^{2},
\label{dege3}  \\  
\frac{e^{2\gamma\phi_{d}}}{\phi_{d}(\phi_{d}+1/\gamma)^{2}}&=&
\gamma Q_{m}^{2}m_{d}^{2}.  
\label{dege4}
\end{eqnarray}

As we mentioned at the begining of this section, $\phi_{d}$ 
is restricted to $\phi_{d}>0$ since the dilaton is 
monotonically decreasing 
outside the horizon for the magnetically charged case. Thus, the left-hand side of 
Eq. (\ref{dege4}) reaches its minimum value $e^{2}\gamma^{3}/4$ 
at $\phi_{d}=1/\gamma$. [Note that $e$ here refers not to electric charge 
but to the Euler number $e=2.71\cdots$.] 

Thus, the solution for $Q_{m}^{2}m_{d}^{2}<e^{2}\gamma^{2}/4$ 
does not have an extreme limit, {\it if $\phi_{d}$ and $\phi_{d}'$ 
have finite values.} To check this, we find the solutions numerically. 
Let us denote the dilaton mass which satisfies 
$Q_{m}^{2}m_{d}^{2}=e^{2}\gamma^{2}/4$ by $m_{d,crit}$. 
For $\gamma =0.5$, $1$ and $\sqrt{3}$, we find that $m_{d,crit}$ takes 
values $6.79\cdots$, 
$13.59\cdots$ and $23.54\cdots$, respectively. 
We pay attention to the qualitative difference depending on the dilaton mass. 

\begin{figure}
\begin{center}
\segmentfig{10cm}{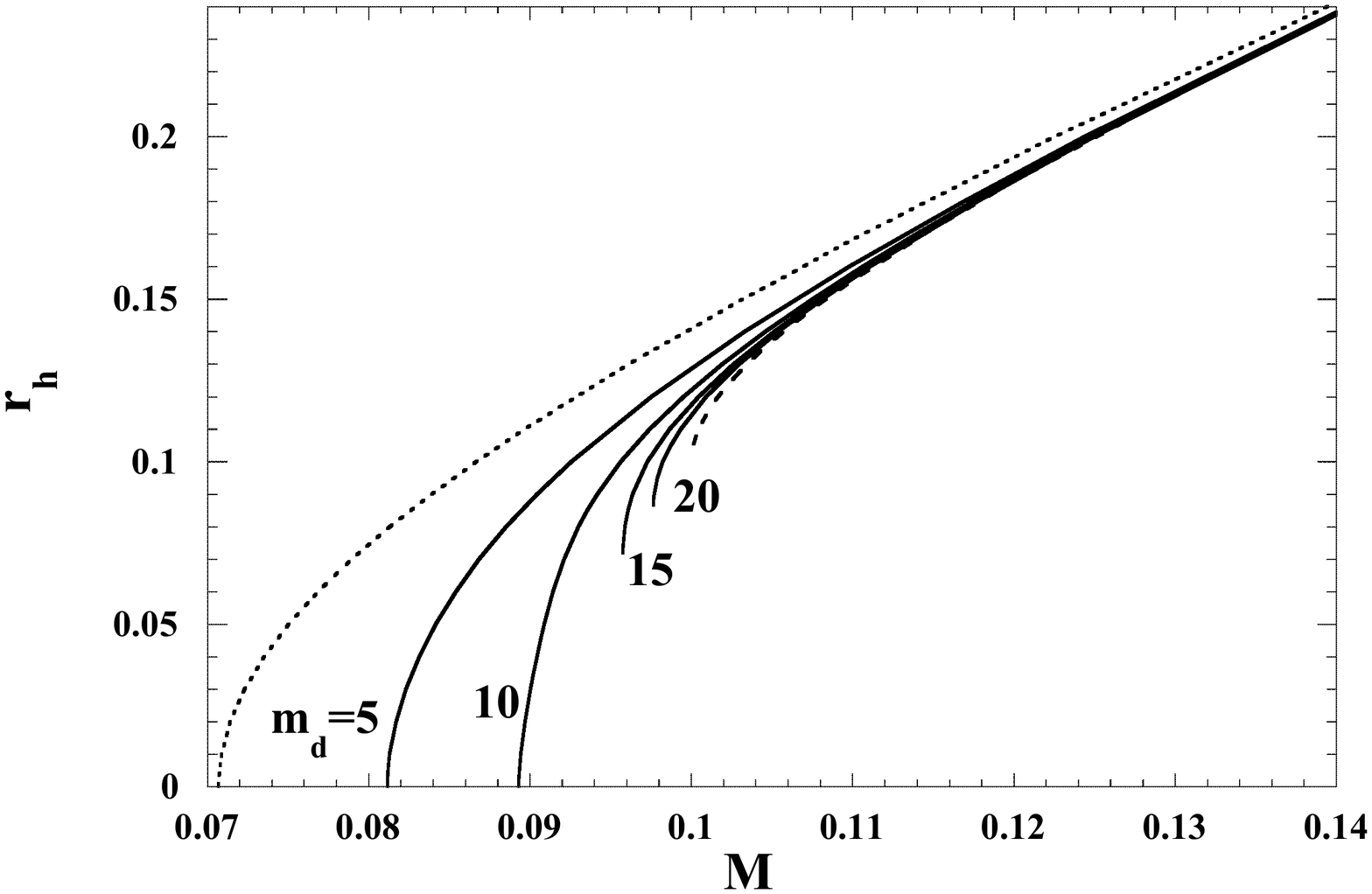}{(a)}  \\
\segmentfig{10cm}{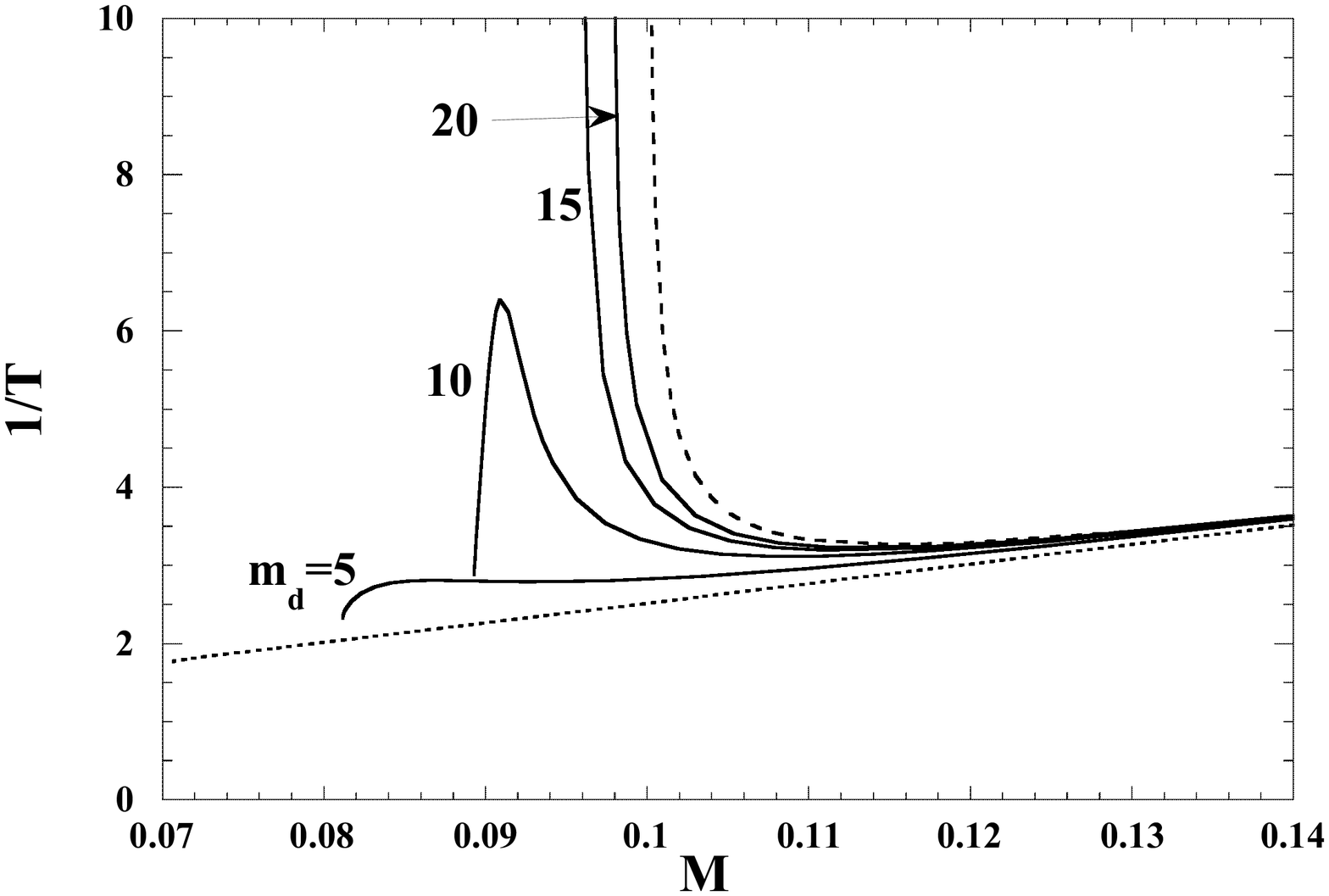}{(b)}  
\caption{(a) The horizon radius $r_{h}$ and (b) the inverse 
temperature $1/T$ in terms of the gravitational mass $M$ with $\gamma =1$. 
The massive solution, GM-GHS solution and RN solution are plotted as 
solid lines, as dotted lines, and a dashed line, respectively. 
\label{Fig2} }
\end{center}
\end{figure}

The horizon radius $r_{h}$ and the inverse temperature $1/T$ are shown 
in terms of the gravitational mass $M$ of the massive dilaton 
black hole for the case $\gamma =1$ in Figs. \ref{Fig2} (a) and (b), 
respectively (solid lines). We also show the GM-GHS and the RN solutions with 
a dotted line and a dashed line, respectively. 

The relationship between the extreme solution and 
the zero temperature is directly connected to the first law of 
black hole thermodynamics. If we fix the magnetic charge, 
it can be expressed as 
\begin{eqnarray}
\frac{dr_{h}}{dM}=\frac{1}{2\pi r_{h}T}\ .
\label{thermo}
\end{eqnarray}
Since $dr_{h}/dM$ approaches infinity in the extreme limit for 
nonzero $r_{h}$ in the RN solution, $T$ approaches zero. 
On the other hand, for the GM-GHS solutions with $\gamma =1$, 
since $dr_{h}/dM$ diverges as $r_{h}$ approaches zero, 
$T$ can remain finite. 

For massive dilaton black holes, our calculation shows that 
there are two distinct cases, depending on the dilaton mass. 
For $m_{d}>m_{d,crit}$, there is an extreme solution 
for nonzero horizon radius. For $m_{d}<m_{d,crit}$, 
the solution continues to exist in the $r_{h}\to 0$ 
limit where $T$ remains finite nonzero value, as in the corresponding 
GM-GHS solution. We defer discussion of the case $m_{d}=m_{d,crit}$ to 
the next section. 

\begin{figure}
\begin{center}
\segmentfig{10cm}{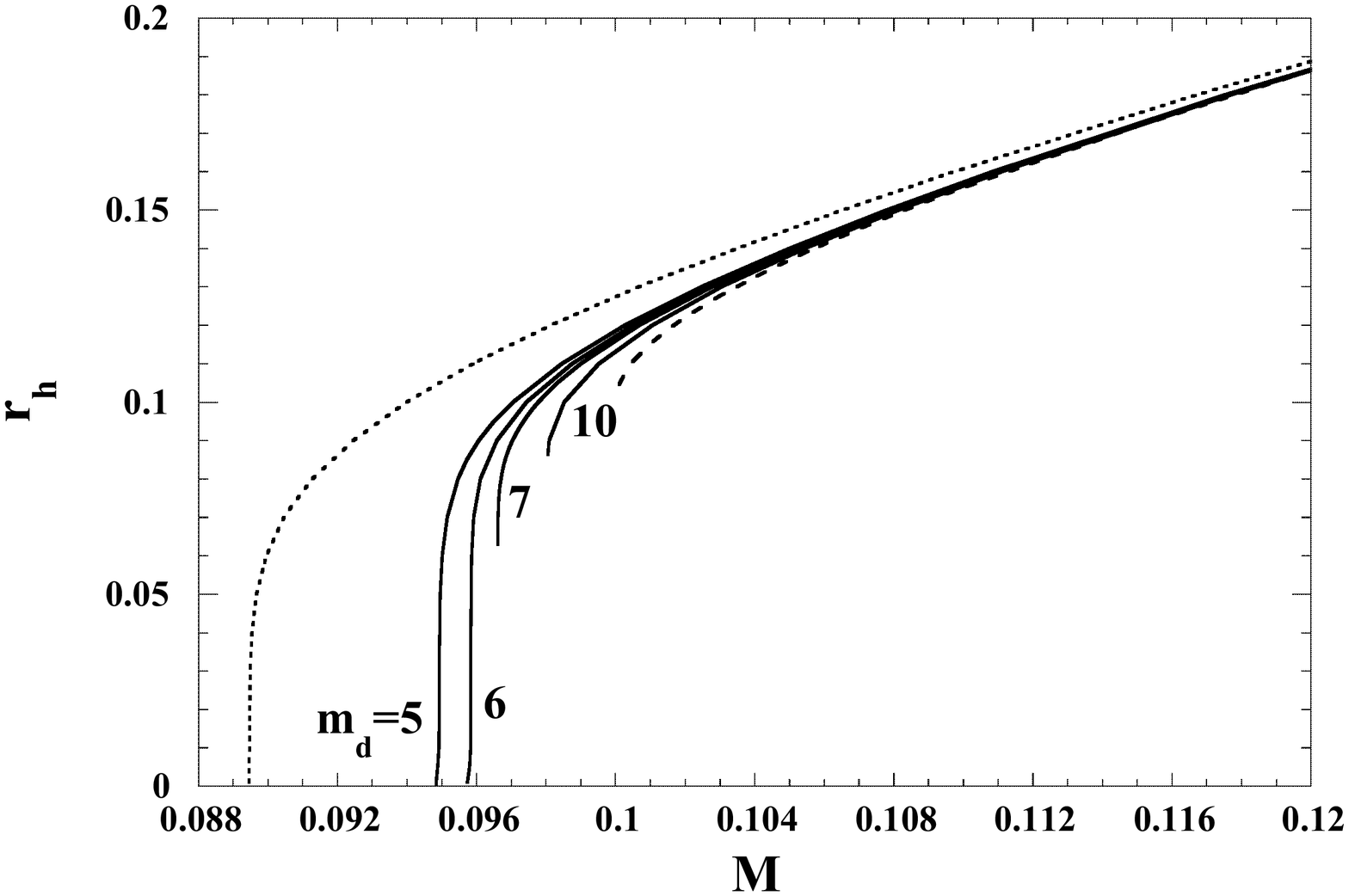}{(a)}  \\
\segmentfig{10cm}{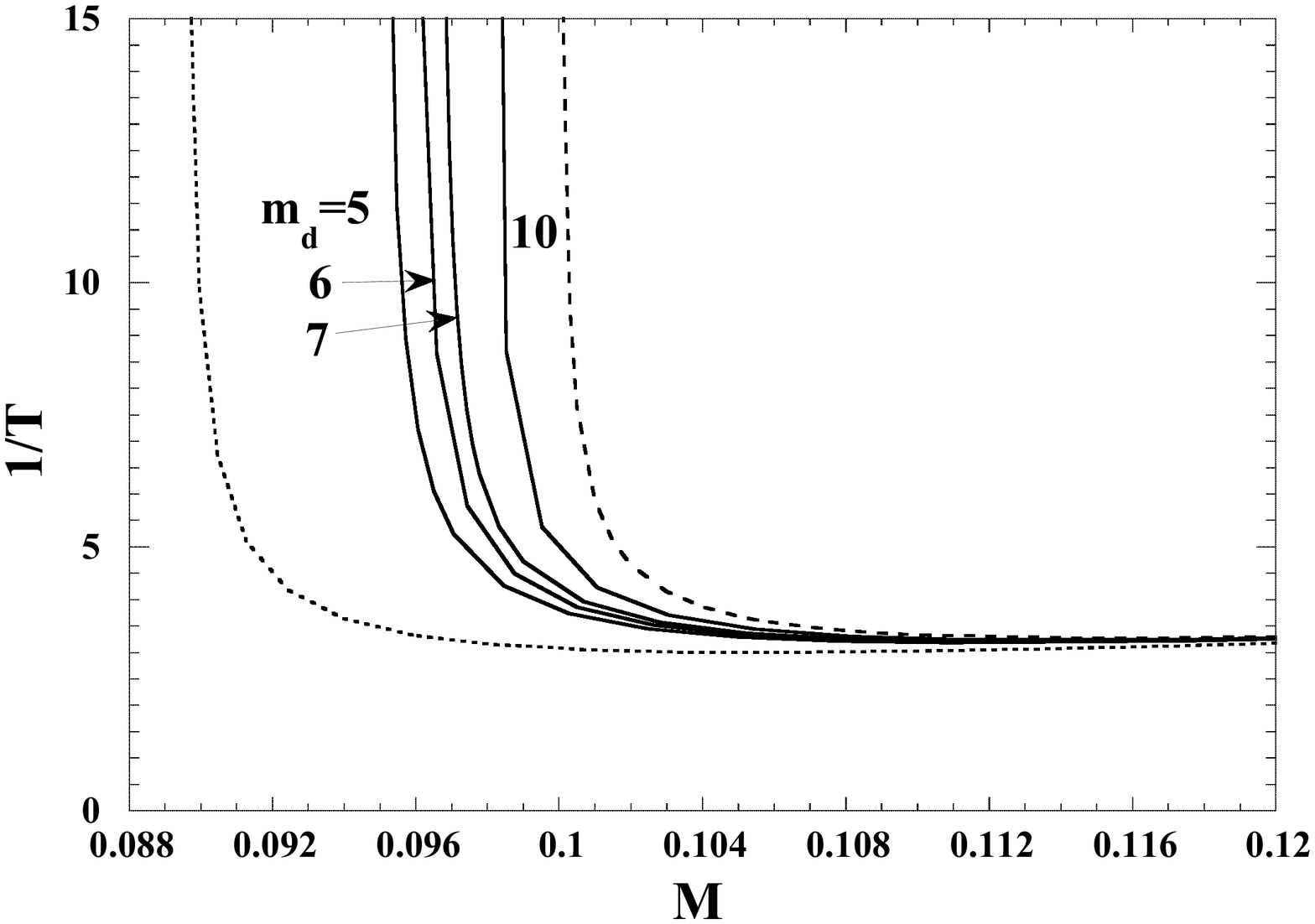}{(b)}  
\caption{(a) The horizon radius $r_{h}$ and (b) the inverse 
temperature $1/T$ in terms of the gravitational mass $M$ with $\gamma =0.5$. 
The massive solution, GM-GHS solution and RN solution are plotted as 
solid lines, as dotted lines, and a dashed line, respectively. 
\label{Fig3} }
\end{center}
\end{figure}

\begin{figure}
\begin{center}
\segmentfig{10cm}{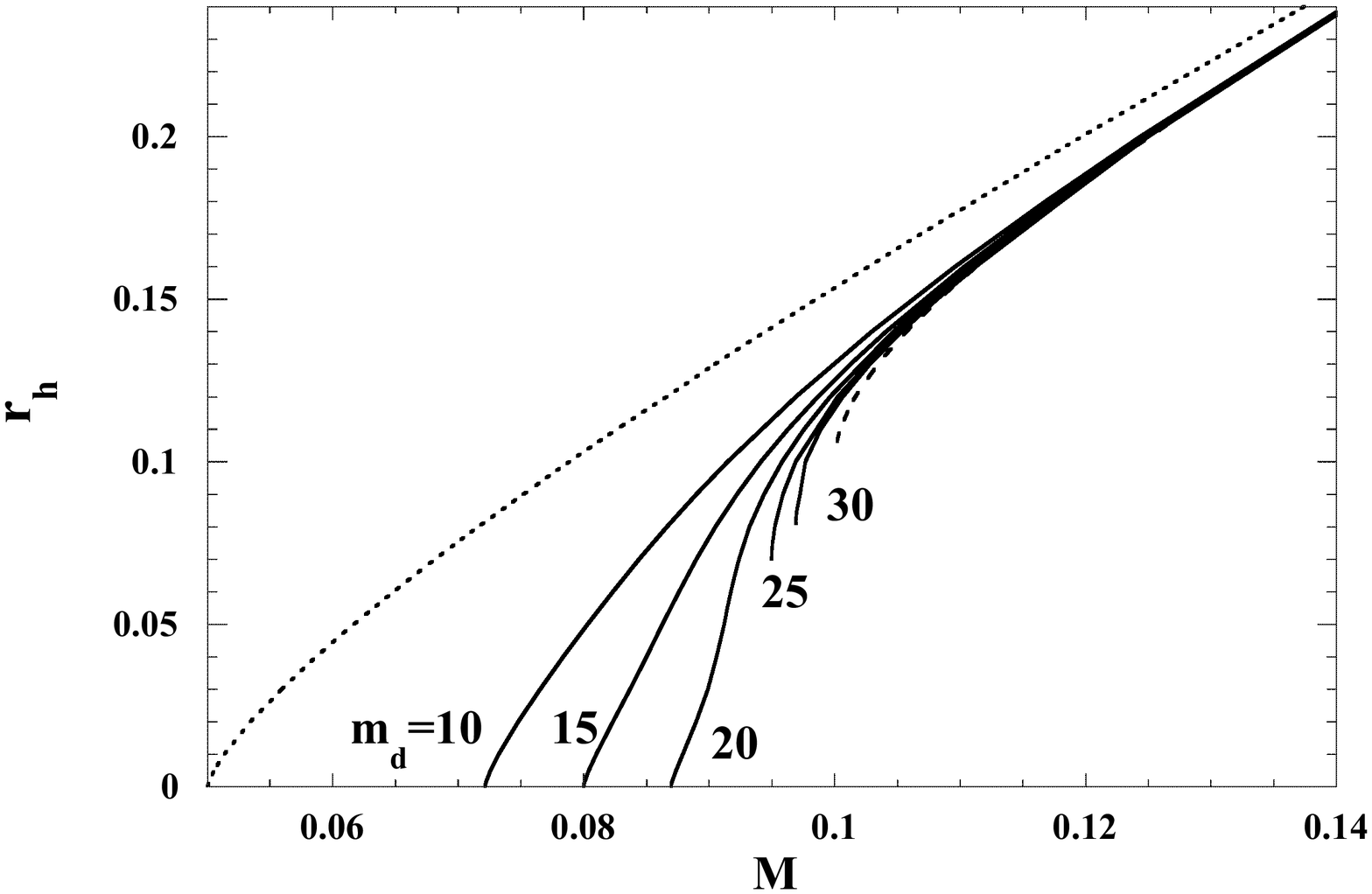}{(a)}  \\
\segmentfig{10cm}{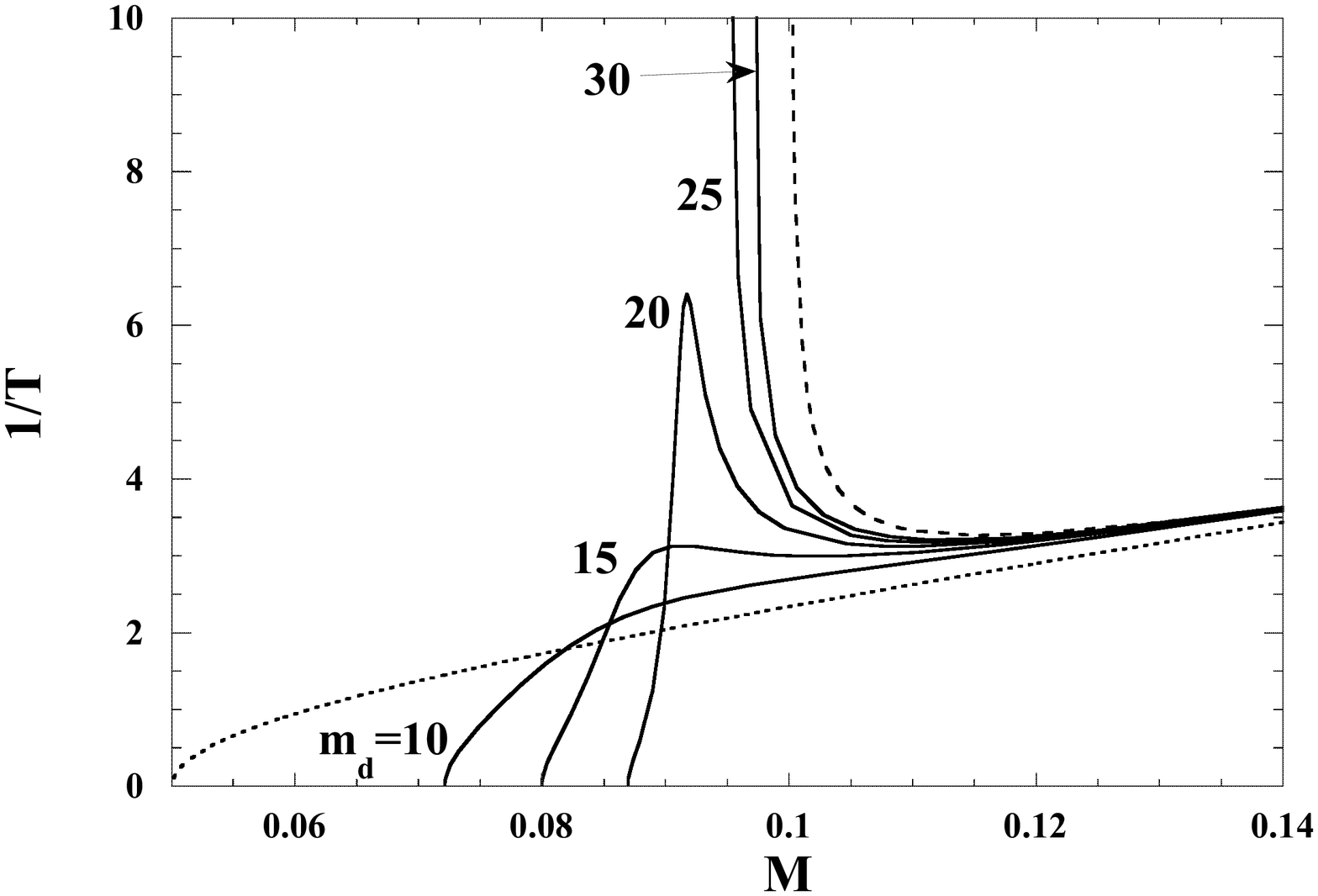}{(b)}  
\caption{(a) The horizon radius $r_{h}$ and (b) the inverse 
temperature $1/T$ in terms of the gravitational mass $M$ with 
$\gamma =\sqrt{3}$. The massive solution, GM-GHS solution and RN solution 
are plotted as solid lines, as dotted lines, and a dashed line, respectively. 
\label{Fig4} }
\end{center}
\end{figure}

We also show the corresponding diagrams for $\gamma =0.5$ and $\sqrt{3}$ 
(suggested by Kaluza-Klein theory) in Figs.~\ref{Fig3} (a) and (b), and 
in Figs.~\ref{Fig4} (a) and (b), 
respectively. Note that the solution in these cases also becomes an extreme 
one for a nonzero horizon radius with $m_{d}>m_{d,crit}$, as in the case 
$\gamma =1$ discussed above. Thus, in some respects these diagrams 
are similar to the case $\gamma =1$. 

It is surprising that the temperature in the zero horizon limit 
for $m_{d}<m_{d,crit}$ closely resembles that of the GM-GHS solution. 
I.e., the temperature diverges, approaches zero or has a finite nonzero value 
in the zero horizon limit, according as $\gamma >1$, $\gamma <1$ or $\gamma =1$, 
respectively. We confirmed this result for various $\gamma$. 
This is also related to the fact that the dilaton field behaves like a 
massless field within its Compton wavelength. However, since the mass term 
of the dilaton field affects the gravitational mass even in the 
zero horizon limit, the solution itself is quite different from the 
GM-GHS solution. Thus, it is not trivial whether or not 
the temperature of a massive dilaton black hole in the zero horizon limit 
resembles that of the GM-GHS solution qualitatively. 

In previous papers\cite{HH}, it was shown that a necessary condition for the 
existence of three horizons is the violation of the strong energy condition 
(SEC). In our example, the violation of the SEC necessarily means that 
$m_{d}>m_{d,crit}$. So, one interesting possibility is that the final fate of the black 
hole depends only on the dilaton coupling for general matter fields if the 
SEC holds. 

For $\gamma \geq 1$, interesting behavior can be expected when 
$m_{d}$ is slightly below $m_{d,crit}$. 
In this case, the temperature approaches zero by reducing the mass. 
However, there is a local maximum of $1/T$ in the $M$-$1/T$ diagram, as 
we can see in Figs.~\ref{Fig2} and \ref{Fig4}. 
So we may find a rapid evaporation below some gravitational mass. 

Related to the above discussion, one may wonder whether the extreme solution 
is a numerical artifact. That is, although the temperature 
approaches zero by reducing the mass, it may have a local maximum in 
the $M$-$1/T$ diagram even in the case $m_{d}<m_{d,crit}$. 
To check this, we should rule out the possibility that we might have stopped 
the calculation before a local maximum in the $M$-$1/T$ diagram 
appeared\cite{foot1}. We do this in the next section.

\section{Near critical dilaton mass}
Here, we show the qualitative difference caused by the dilaton mass 
in a more explicit way. We do this by means of a diagram for the case 
$\gamma =1$. However, our discussion is general for arbitrary $\gamma$. 

\begin{figure}[htbp]
\singlefig{10cm}{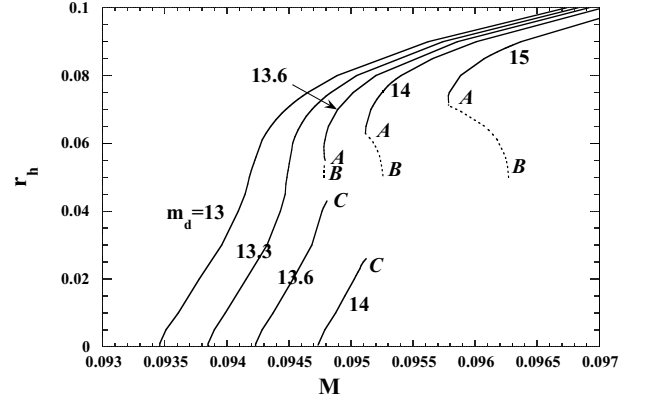}{}  
\caption{The horizon radius $r_{h}$ as a function of the gravitational 
mass $M$ with $\gamma =1$. 
\label{Fig6} }
\end{figure}

In Sec. IV, we noted that there is an extreme solution for 
$m_{d}>m_{d,crit}$. This becomes more clear if the existence of 
the inner horizon and its merger with the event horizon in the extreme limit 
can be demonstrated. We plot the event horizon $r_{h}$ as a function of the 
gravitational mass $M$ of the massive dilaton black hole for 
$\gamma =1$ near $m_{d,crit}$ in Fig.~\ref{Fig6} (solid lines). 
We also show the ``inner horizon" with dotted lines. 

As was shown in previous papers\cite{HH}, there is no ``inner horizon" for 
large black holes, while one does appear for small black holes. This kind of 
behavior is also found in charged black holes with 
Born-Infeld type nonlinear electrodynamics\cite{Demianski}. It is also 
worth mentioning that the inner horizon disappears in the Born-Infeld type 
black hole if we include a dilaton field\cite{Tamaki2}. 

Since we need fine-tuning of the 
dilaton field $\phi$ to satisfy regularity at the inner 
horizon, a curvature singularity appears there in general and 
is spacelike as in the GM-GHS solution. 

The ``inner horizon" disappears below the point $B$ in Fig.~\ref{Fig6} 
where $dr_{h}/dM\to \infty$ and $r_{h}\sim 1/m_{d}$. 
The point $A$ corresponds to the point where two horizons separate. 
Thus, this diagram supports our contention that there is an extreme 
solution for $m_{d}>m_{d,crit}$. 

Thus, it is astonishing that there are solutions below points $C$ 
for $14.8\agt m_{d}>m_{d,crit}$. In our calculation, gravitational mass 
of the point $C$ coincides with that of the point $A$. So it is supposed that 
uniqueness of solutions for fixed parameters holds in our system. 
Temperature is finite and the SEC holds in these cases. The reason why 
solutions below points $C$ exist for $14.8\agt m_{d}>m_{d,crit}$ is 
under investigation. 

As $m_{d}$ approaches $m_{d,crit}$ for 
$m_{d}>m_{d,crit}$, the curve between $A$ and $B$ becomes 
small. For $m_{d}=13.6$, $A$ almost coincides with $B$ and $C$. 
It is natural to suppose that $A$ coincides with $B$ and $C$ for 
$m_{d}=m_{d,crit}$. This is related to the existence of a 
triply-degenerate horizon, as pointed out in previous papers\cite{HH}. 

Finally, we investigate the properties of $\phi_{h}$ for various horizon 
radii, since $\phi_{h}$ cannot be determined analytically, and whether or 
not this quantity diverges is the key to the existence 
of the extreme solution, as discussed in Sec.IV. 

Let us first examine the condition on $\phi_{h}$ for the existence of the 
regular horizon. From Eqs. (\ref{eqn:einstein}) and (\ref{eqn:dilaton}), 
we obtain 
\begin{eqnarray}
\frac{r_{h}^{4}m_{d}^{2}}{Q_{m}^{2}}&<&\frac{\gamma}{\phi_{h}}
e^{-2\gamma \phi_{h}},
\label{cond1}  \\  
\frac{r_{h}^{4}m_{d}^{2}}{Q_{m}^{2}}&<&
\frac{1}{\phi_{h}^{2}}\left(\frac{r_{h}^{2}}{Q_{m}^{2}}-
e^{-2\gamma \phi_{h}}\right)  
\label{cond2}
\end{eqnarray}
by requiring that $m_{h}'<1/2$ and $\phi_{h}'<0$, respectively. 
$\phi_{h}'<0$ is required if $\phi$ is to decrease monotonically. 
We find from Eq. (\ref{cond1}) that $\phi_{h}r_{h}$ must converge 
to zero for the existence of the $r_{h}\to 0$ limit. 

\begin{figure}[htbp]
\singlefig{10cm}{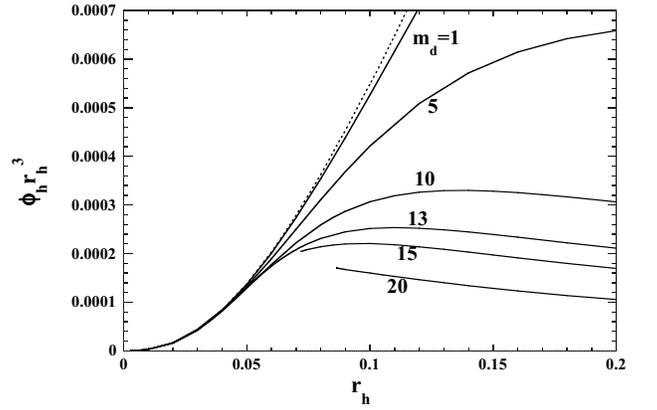}{}
\caption{The relationship between $r_{h}$ and $\phi_{h}r_{h}^{3}$ 
for $\gamma =1$ (shown also for the GM-GHS solution as a dotted line.). 
\label{Fig7}}
\end{figure}

A plot of $r_{h}$-$\phi_{h}r_{h}^{3}$ for $\gamma =1$ is shown 
in Fig.~\ref{Fig7}. We find that $\phi_{h}r_{h}^{3}$ increases as 
$r_{h}$ decreases when the horizon is large enough to satisfy 
$r_{h}>1/m_{d}$. 

On the other hand, when the horizon becomes too 
small to satisfy $r_{h}<1/m_{d}$, then $\phi_{h}r_{h}^{3}$ 
decreases as $r_{h}$ decreases. Note that the curves for $m_{d}<m_{d,crit}$ 
approach those of the GM-GHS solution as $r_{h}$ decreases. 

For $m_{d}=20$, we can safely conclude that, since 
$\phi_{h}r_{h}^{3}$ monotonically increases but does not diverge 
as $r_{h}$ decreases, $m_{h}'=1/2$ is satisfied for a nonzero horizon radius. 
For $m_{d}=15$, no conclusion can be drawn from this diagram. 

\begin{figure}[htbp]
\singlefig{10cm}{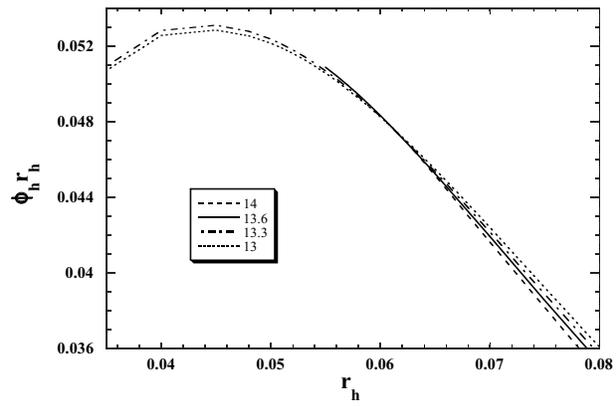}{}
\caption{The relationship between $r_{h}$ and $\phi_{h}r_{h}$ 
for $\gamma =1$. 
\label{Fig8}}
\end{figure}

To judge whether or not the extreme solution is realized exactly for nonzero 
horizon radius, we need to investigate the behavior of $\phi_{h}r_{h}$ for 
dilaton masses near $m_{d,crit}$ ($m_{d}=13\sim 14$) as shown in 
Fig.~\ref{Fig8}. Since $\phi_{h}r_{h}$ increases monotonically as $r_{h}$ 
decreases for $m_{d}=13.6$ and $14$, we can conclude that 
$m_{h}'=1/2$ is satisfied for a nonzero horizon radius. 
Thus, the black hole properties change at $m_{d,crit}$, as shown by 
the behavior of $\phi_{h}$. 

\section{Conclusion and Discussion}
We have investigated static spherically symmetric solutions in the 
Einstein-Maxwell-dilaton system with a dilaton potential $V=m_{d}^{2}\phi^{2}$. 
We considered the magnetically charged solution for various values of the 
dilaton coupling. We confirmed that the properties of black holes change 
qualitatively at $m_{d,crit}$. 

For $m_{d}\geq m_{d,crit}$, which means the violation of the SEC, 
there is an extreme solution for nonzero horizon radius. 
For $m_{d}<m_{d,crit}$, the solution continues to exist in the $r_{h}\to 0$ 
limit where the temperature remains either finite and nonzero, diverges, or 
approaches zero for $\gamma =1$, $\gamma >1$ and $\gamma <1$, respectively. 

This property may also hold when the SEC is not violated. For example, the 
Einstein-Yang-Mills-dilaton (EYMD) system has been considered 
before\cite{SU2EYMD,Kunz,Inada} and exhibits thermodynamic properties similar 
to those above, if the black hole has a globally magnetic (or electric) charge. 
To prove this for general matter fields, or to find a 
counterexample is a task we hope to take up in future research. 

\section*{Acknowledgments}
Special thanks to Kei-ichi Maeda and Takashi Torii for useful 
discussions. This work was supported by the Waseda University Grant 
for Special Research Projects.




\begin{thebibliography}{99}
\bibitem{foot2}
For the notation, see our action (\ref{eqn:action}).
\bibitem{foot1}
Actually in \protect\cite{TY}, we misjudged some cases which have a solution 
in the $r_{h}\to 0$ limit as cases in which there is an extreme solution.
\end{thebibliography}
\end{document}